\documentclass[10pt, conference, letterpaper]{IEEEtran}
\IEEEoverridecommandlockouts
\usepackage[utf8]{inputenc}
\usepackage[T1]{fontenc}
\usepackage{graphicx}
\usepackage{epsfig} 
\usepackage[bottom]{footmisc}
\usepackage{cite}
\usepackage{amsmath, amssymb, amsfonts}
\usepackage{fancyhdr}
\usepackage{float}
\usepackage{algorithm}
\usepackage{algpseudocode} 
\usepackage{algorithmicx}
\usepackage{textcomp}
\usepackage{multirow}
\usepackage[xcdraw, table]{xcolor}
\usepackage{booktabs} 
\usepackage{caption}
\usepackage{subfigure}
\usepackage{rotating}
\usepackage{tablefootnote}
\usepackage{tikz}
\usepackage{mathtools}
\usepackage{ctable} 
\def\BibTeX{{\rm B\kern-.05em{\sc i\kern-.025em b}\kern-.08em
    T\kern-.1667em\lower.7ex\hbox{E}\kern-.125emX}}

\begin{document}

\title{\huge CADRE: Card-Agnostic Domain-Aligned RF Embeddings for Virtual PIN Pads on Passive NFC Cards 
}
\author{
\IEEEauthorblockN{Dickson Akuoko Sarpong and Hongzhi Guo}
\IEEEauthorblockA{School of Computing, University of Nebraska--Lincoln, Lincoln, NE, USA \\
Email: dakuokosarpong2@huskers.unl.edu; hguo10@unl.edu}
}

\maketitle

\begin{abstract}
Near Field Communication (NFC) cards are widely used for identification, but their passive nature often limits the ability to incorporate additional security mechanisms. As a result, anyone holding the card may be incorrectly recognized as an authenticated user. To overcome this limitation, this paper presents a secure manual password input framework using a virtual PIN pad for passive NFC cards. Users input passwords by pressing designated regions on the card, which induces measurable impedance variations in the NFC antenna. These variations change the RF signals subtly, and a deep learning model is used to infer the intended password from the resulting signal patterns. A key challenge is that identical press interactions can produce significantly different responses across NFC cards, which yields unreliable recognition. To address this, we introduce a lightweight recognition approach that operates directly within the RF feature space at the penultimate layer of a temporal neural encoder. An adversarial domain-alignment module reshapes virtual PIN pad press-response embeddings into compact, card-invariant clusters, which enables stable and consistent recognition across heterogeneous cards. To support model training and evaluation, a reconfigurable software-defined radio (SDR) testbed is developed, and PIN pad press-response data are collected from commercially available ISO/IEC 15693 cards. Recognition is performed using a Mahalanobis distance metric derived from a calibration-based covariance model that captures feature correlations. Experimental results show that the proposed system achieves a 98.20\% recognition acceptance rate and remains robust under substantial noise degradation. The framework is fully card-agnostic and can be seamlessly integrated into existing NFC infrastructures.
\end{abstract}


\IEEEpeerreviewmaketitle


\section{Introduction}
\label{intro}

Near Field Communication (NFC) has become an essential technology in modern access control, payment systems, and device-to-device interaction. Its extensive adoption is driven by its low cost, ease of use, short-range secure operation, and the use of standardized ISO/IEC communication protocols \cite{Wang2025IDCCIC,Sarpong2025ModelAgnosticUQ}. Although digital authentication provides convenience, it also introduces security vulnerabilities. For instance, anyone in possession of an NFC card may be incorrectly recognized as an authenticated user. Furthermore, in relay and protocol forwarding attacks, adversaries can clone or proxy NFC cards without physically owning them \cite{Yang2025EnhancedES,Teng2025DynamicPF}. These attacks are possible because the reader’s decision-making process relies solely on digital card information.

Recent work has shown that the physical-layer characteristics of NFC interactions, such as coil geometry, resonance behavior, and mutual inductive coupling, carry rich analog signatures that are inherently difficult to synthesize, forward, or clone \cite{Sarpong2025ModelAgnosticUQ}. RF fingerprinting leverages these device-specific features for identification, and prior results demonstrate its potential for mitigating card cloning attacks \cite{Adigopula2024DCNNRFFNFCAN}. In addition, practical NFC systems operate robustly over a frequency band (approximately 13-14 MHz) rather than at a single monotone frequency, which provides diverse RF features that further motivate incorporating analog effects into security mechanisms \cite{Yang2024JumpOO}. However, RF fingerprinting can only authenticate the NFC card itself, not the user who possesses it.

A common solution for authenticating both the user and the NFC card is to employ an NFC reader equipped with PIN pads and displays to allow password input after the card is tapped. However, this approach increases the cost and complexity of the reader. Additionally, in certain scenarios, such as during a pandemic, users may be unwilling to touch shared PIN pads. Prior work has also shown that passwords entered on reader-side PIN pads can be inferred using thermal cameras or pre-deployed recording devices. This further compromises NFC security. Meanwhile, NFC cards themselves are passive devices without batteries or built-in interfaces for password input. Consequently, enabling both NFC card and user authentication with low-cost, simple NFC readers and fully passive NFC cards remains a significant challenge.

\begin{figure}[t]
    \centering
    \includegraphics[width=0.9\columnwidth]{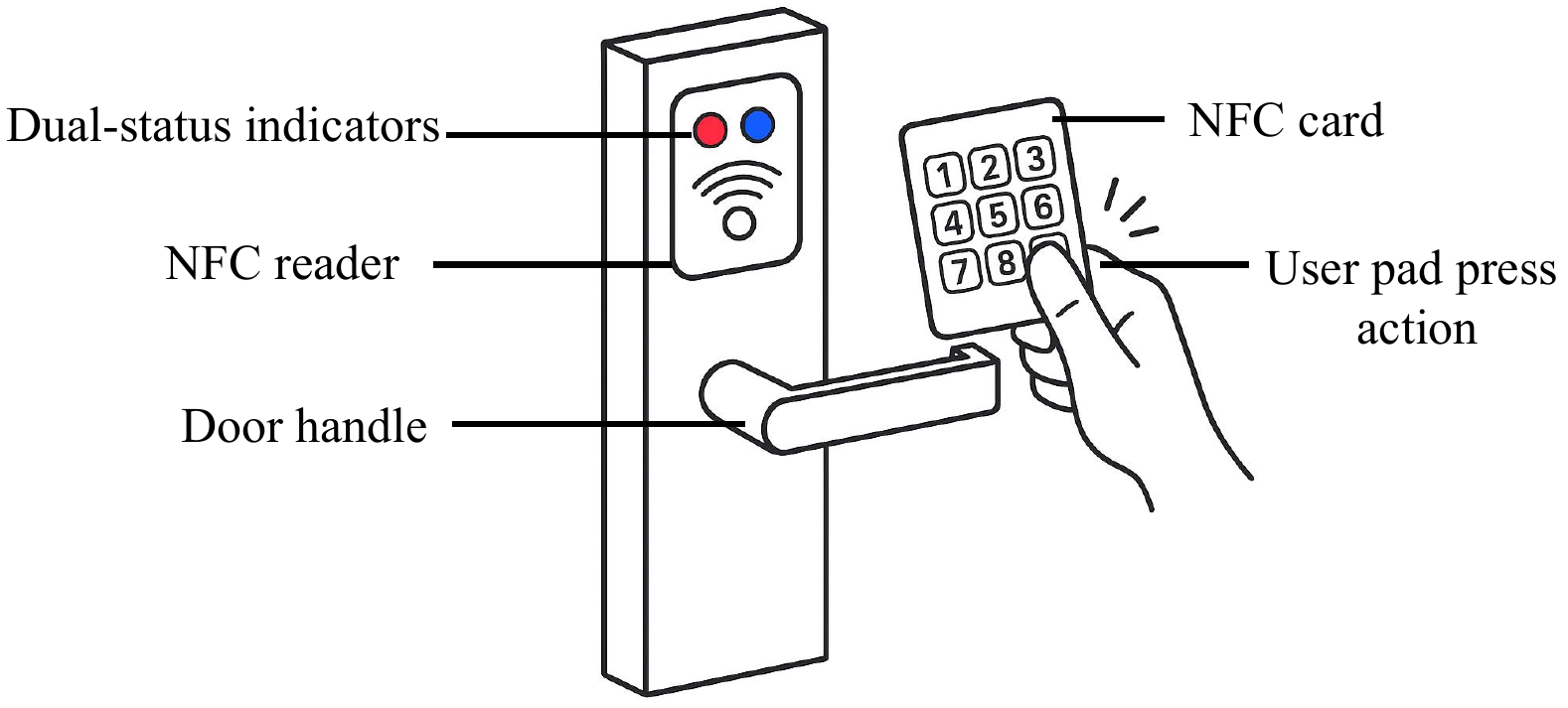}
    \caption{Application example of the \emph{virtual PIN pad} in NFC door mounted access system.} 
    \label{fig:pinpad_apply}
    \vspace{-13pt}
\end{figure}

In this work, we introduce a low-cost physical-layer authentication mechanism that can be seamlessly integrated into existing NFC systems. Rather than relying solely on digital identification data, we enable NFC users to input passwords using passive NFC cards, which turns the card into a \emph{personal virtual PIN pad}. Multiple predefined regions on the card are marked with PIN digits, as illustrated in Fig.~\ref{fig:pinpad_apply}. During operation, the user places a small passive resonant coil, such as an NFC keyfob, onto one of these designated regions, which serves as a virtual press. Each press perturbs the magnetic coupling between the reader antenna and the card antenna, and produces a distinct and repeatable signature in the reader’s received signals. By detecting the small coil's location on an NFC card, we can recognize the PIN digit that the location represents. Small coil placements at different regions thus function as PIN inputs. In this way, we can authenticate both the NFC card and the user. 

In the target application, the NFC reader provides two simple status indicators, as shown in Fig.~\ref{fig:pinpad_apply}. A blue light indicates that a standard NFC tap is sufficient, whereas a red light signals that the system requires a short sequence of coil placements analogous to entering PIN digits. The user holds the card near the reader and performs the required virtual presses at the predefined locations. For each press, the reader captures a short signal sample and determines whether to accept or reject the input.

Developing a robust PIN digit recognition framework on passive NFC cards poses several challenges. 
First, the wireless signals produced by a press are short high-frequency signals whose subtle differences are easily obscured by hardware-dependent distortions. Extracting stable, interaction-specific features therefore requires obtaining a representation that suppresses device artifacts while preserving the virtual PIN digits identity. 
Second, NFC hardware exhibits significant domain shift across different cards and readers. Variations in coil geometry, Q-factor, parasitics, shielding, and carrier detuning cause signals corresponding to the same PIN digit to appear differently across devices \cite{Fischbacher2024CommunicationAP}. As a result, signals used for recognition on one card often fail to generalize to others. 
Third, any added authentication mechanism must remain fully compatible with existing ISO/IEC NFC protocols and require no power from the card. 
Finally, a reliable recognition framework must exhibit stable acceptance and rejection behavior even when evaluating samples from previously unseen cards, readers, or environmental conditions. 
These challenges motivate the need for a card-invariant PIN digit press-recognition system for personal virtual PIN pad on an NFC card.

In this paper, we present CADRE, a Card-Agnostic Domain-Aligned RF Embedding framework for NFC PIN digit recognition and authentication. This is the first system using passive NFC cards as PIN input devices. The system is entirely passive, requires no modifications to NFC protocols, and operates using only short sequences of signals. The major contributions of this work are summarized as follows:
\begin{itemize}
    \item First, we develop an analytical model of the multi-coil NFC PIN pad system that captures reader coil, card coil and PIN button coil couplings and interactions. We further validate the model through Finite Element Method (FEM) simulations. Together, they demonstrate the feasibility of the proposed virtual PIN pad concept.
    \item Second, we construct a fully reconfigurable ISO/IEC 15693-compliant experimental testbed using software-defined radios (SDRs), which supports NFC reader signal transmission, card signal reception, and systematic data collection across diverse cards.
    \item Third, we design a temporal neural encoder together with a domain adaptation pipeline that learn card-invariant PIN digit representations and supports stable classification across heterogeneous NFC cards.    
    \item Finally, we introduce a Mahalanobis multivariate distance-based PIN digit recognition mechanism that leverages calibration-driven covariance geometry to provide reliable acceptance and rejection decisions under cross-card domain shift, including for previously unseen cards.
\end{itemize}
Although CADRE generalizes across different NFC cards, in this paper we assume that the cards are from the same model and the same vendor. Cross-vendor or different types of NFC cards are not considered in this paper.

The remainder of the paper is organized as follows. Section~\ref{system_model} presents the NFC-based virtual PIN pad design along with its electromagnetic (EM) and circuit-level modeling. Section~\ref{Data collect} describes the prototype design and data acquisition methodology. It also details the domain-adversarial training framework for learning card-invariant PIN digit embeddings and introduces the Mahalanobis distance-based PIN digit recognition scheme. Section~\ref{Results} reports experimental results and key findings. Finally, Section~\ref{Conclude} concludes the paper.


\section{Theoretical and Simulation Model}
\label{system_model}
In this section, we present the NFC-based virtual PIN pad system and its underlying EM and circuit-level model, which together provide the theoretical foundation for CADRE. The system comprises a reader, an NFC card, and a small button coil. The button coil can be placed at different predefined regions on the card to represent PIN digits. For modeling and analysis, we characterize how region-dependent reader-card-coil coupling affects the original reader-card interaction and introduces distinct features that enable reliable PIN digit recognition. In other words, a PIN digit is represented by a specific location of the button coil on the passive NFC card.

\subsection{Near-Field Magnetic Coupling Model}

\begin{figure}[t] 
    \centering
    \includegraphics[width=0.93\columnwidth]{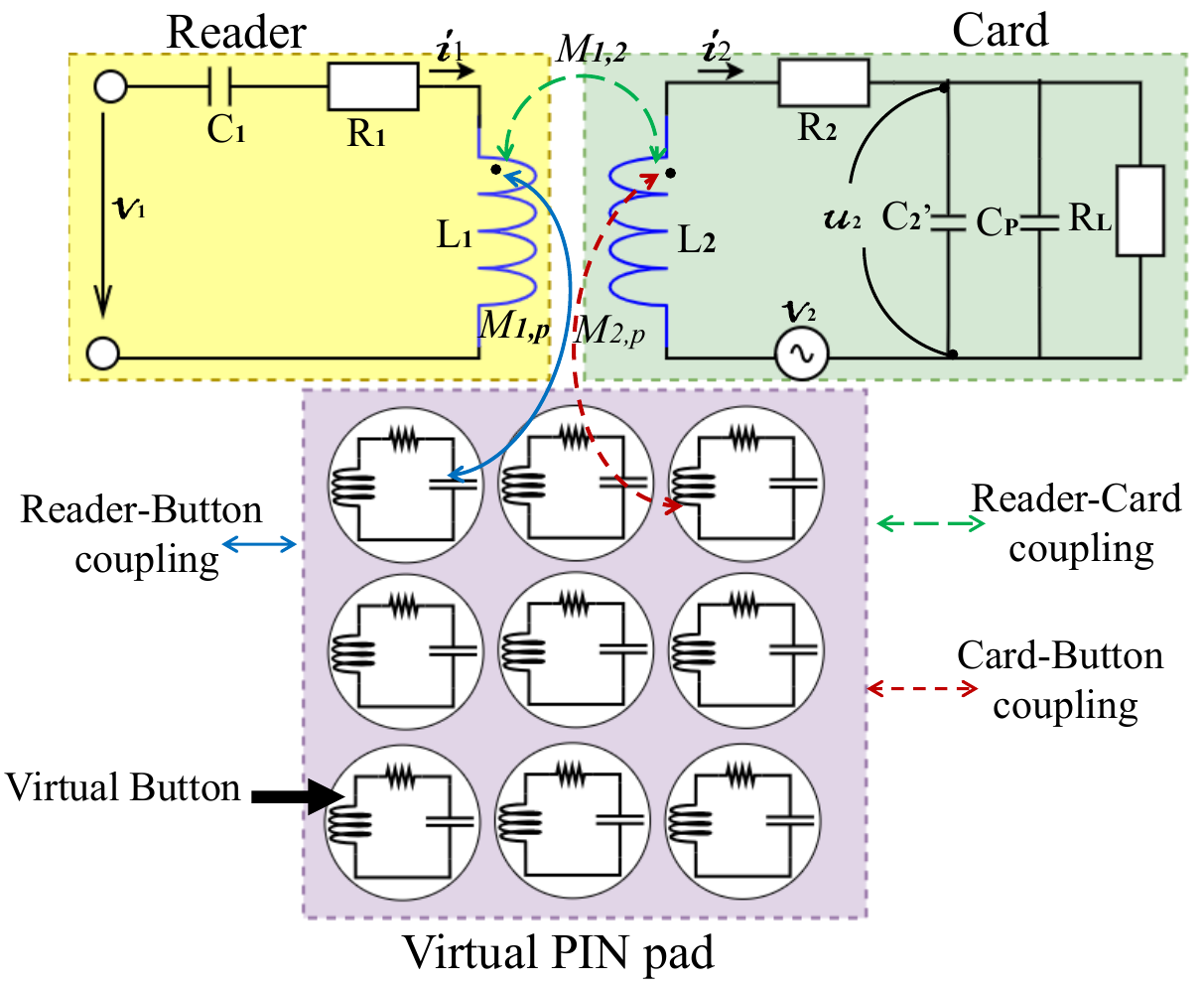}
    \caption{A simplified circuit diagram for the NFC reader-card system with a virtual PIN pad on an NFC card.}
    \label{fig:sys_model}
    \vspace{-15pt}
\end{figure}

To study how the button coil affects the original interaction between the reader and card coils, we develop an analytical model. The button coil is placed sufficiently close to the card so that the interaction occurs in the near-field regime. To ensure the repeatability, a card holder with predefined button position can be used. Each coil is modeled as a planar multi-turn circular loop with radius $a$ and number of turns $n$ in a homogeneous medium. The mutual inductance $M_{iq}$ between the $i$-th and $q$-th coils, with radii $a_i$ and $a_q$, number of turns $n_i$ and $n_q$, vertical separation $|z_i - z_q|$, and lateral offset $\lvert p \rvert$, is given by \cite{conway2007inductance}:
\begin{equation}
M_{iq} = \mu_0 \pi n_i n_q a_i a_q \int_0^\infty J_0(sp) J_1(s a_i)J_1(s a_q)\, e^{-s |z_i - z_q|} ds,
\end{equation}
where $\mu_0$ is the permeability of free space, $s$ is the spectral integration variable, and $J_0(\cdot)$ and $J_1(\cdot)$ are Bessel functions of the first kind. This expression is derived from the axial component of the vector potential and remains valid even for closely spaced or strongly coupled coils, where classical dipole-based approximations become inaccurate. The self-inductance $L_i$ of an individual coil is obtained by taking the special case $i = q$ and $p = 0$, yielding
\begin{equation}
L_i = \mu_0 \pi n_i^2 a_i^2 \int_0^\infty \left[J_1(s a_i)\right]^2 ds.
\end{equation}

As shown in Fig.~\ref{fig:sys_model}, the reader is modeled as a resonant RLC circuit. The reader antenna, i.e., coil~1 is modeled using a self-inductance $L_1$ with associated series resistance $R_1$. To ensure optimal receive power and enable continuous wave (CW) generation, the coil is configured as part of a series resonant circuit by introducing a capacitor $C_1$. This configuration maximizes the current required to effectively activate the card and maintain CW excitation for card interrogation. The total impedance of the reader is
$
Z_1 = R_1 + j\omega L_1 + \frac{1}{j\omega C_1},
$
where $\omega$ is the angular excitation frequency. The reader circuit is tuned such that its excitation frequency coincides with the resonance frequency $f_c$. At this point, the inductive and capacitive reactances cancel and reduce the impedance to a purely resistive form. The reader current $i_1$ is maximized at this resonant frequency and is given by
$
i_1 = {v_1}/{R_1}, 
$
where $v_1$ is the reader’s drive voltage.

The card antenna is modeled using a parallel RLC network composed of a self-inductance $L_2$, resistance $R_2$, capacitance $C_2'$, and an auxiliary parallel capacitor $C_P$ to enable load modulation, as shown in Fig.~\ref{fig:sys_model}. In typical NFC operation, the card IC controls a load resistor $R_L$ through a switching mechanism. When the switch is closed, i.e., modulation off, the capacitor $C_2'$ remains connected to the antenna, resulting in a higher effective load impedance and a correspondingly weaker reflected signal. When the switch is open, i.e., modulation on, $C_2'$ is disconnected, which lowers the load impedance and produces a stronger reflected response. This impedance variation encodes information in the response signals. The card impedance is given by
\begin{equation}
Z_2 = R_2 + j\omega L_2 + Z_2',
\end{equation}
where $Z_2' = \left( \frac{1}{j\omega C_2} \parallel R_L \right)$ is the equivalent load impedance due to the card ICs and $C_2 = C_2' + C_P$. After harvesting energy from the CW signal, the card experiences an induced voltage $v_2 = j \omega M_{12} i_1$ due to its mutual coupling with the reader coil, where $M_{12}$ represents the coupling between reader and the card. This voltage drives a current,
$i_2 = {j \omega M_{12} i_1}/{Z_2}.$
The resulting current modulates the card’s response, which is reflected back to the reader and perceived as a change in its input impedance.

As illustrated in Fig.~\ref{fig:sys_model}, the virtual PIN pad is realized by predefining discrete button locations on the NFC card surface, each corresponding to a press region for the button coil. The button coil, i.e, coil $p$ is a passive RLC resonator with coil self-inductance $L_{p}$, coil resistance $R_{p}$, and a capacitor $C_{p}$. When the user presses a predefined region on the card shown in Fig.~2 as virtual button with a total of $N=9$ predefined virtual buttons, a button coil is positioned at that location and forms a local RLC branch that magnetically couples to both the reader antenna and the card antenna.

The button coil is a passive magnetic coupler that perturbs the original reader–card antenna interaction. The impedance of the button coil is given by
$Z_{p} = R_{p} + j\omega L_{p} + \frac{1}{j\omega C_{p}}.$
For any press on virtual button on card, the coupled system consists of three coils: the reader, the card, and the button coil. The excitation originates from the reader and the card and button coils remain passive. According to Kirchhoff’s law, the voltages and currents across all coupled coils satisfy
$\mathbf{v} = \mathbf{Z}\mathbf{i},$
which results in the following matrix representation for the reader-card-button coil system:
\begin{equation}
\begin{bmatrix}
v_1\\[2pt]
v_2\\[2pt]
v_{p}
\end{bmatrix}
=
\begin{bmatrix}
Z_{1}          & j\omega M_{12}   & j\omega M_{1p} \\
j\omega M_{12} & Z_{2}            & j\omega M_{2p} \\
j\omega M_{1p} & j\omega M_{2p}   & Z_{p}
\end{bmatrix}
\begin{bmatrix}
i_1\\[2pt]
i_2\\[2pt]
i_{p}
\end{bmatrix},
\label{eq:Zmatrix_single}
\end{equation}
where $\mathbf{v} \in \mathbb{C}^{3\times 1}$ and $\mathbf{i} \in \mathbb{C}^{3\times 1}$ denote the vectors of complex voltages and currents of the three coupled coils, respectively. Specifically, $v_p$ represent the induced voltages in the passive button coil, while $i_p$ denote the corresponding button coil current. The mutual inductances $M_{1p}$ and $M_{2p}$ in Eq.~\eqref{eq:Zmatrix_single} are location dependent and capture the effect of pressing different predefined virtual buttons on the card.

\subsection{Received Signals Affected by the PIN Button Coil}
During interrogation, the reader generates a current $i_1$ that induces voltages in the card antenna and in the button coil placed at a predefined press region. 
The card alternates between two impedance states controlled by a binary switching function $u(t)\in\{0,1\}$, which governs the load modulation specified in ISO/IEC~15693 \cite{10.5555/861917}. By EM reciprocity, the currents induced in any coupled coil contribute a reflected impedance at the reader, which can be expressed as \cite{Guo2015M2ICF}
\begin{equation}
Z_{\mathrm{ref}}=
\frac{\omega^{2} M_{12}^{2}}{Z_{2}}+\frac{\omega^{2} M_{1p}^{2}}{Z_{p}},
\end{equation}
where the denominators are determined jointly by the card load-modulation state and the presence of the button coil at the pressed region. Different PIN digits affect $M_{1p}$ which results in subtle change of reader received signals.

The received response signal is shaped by the frequency response $H(f)$ of the reader's RF front end, which includes both the antenna and its associated circuitry. In addition, the strength of the reflected signal depends on the effective coupling factor $k$, which is $M_{1p}/\sqrt{L_1L_p}$, when the button coil is placed at a particular virtual button location. Combining these effects, the measured baseband spectrum can be expressed as
\begin{equation}
S(f) = H(f)\!\cdot\! k \!\cdot\!
\left[A_{\mathrm{CW}}\delta(f) + A_{\mathrm{card}}\bigl(\delta(f - f_s) + \delta(f + f_s)\bigr) \right].
\label{eq:signal}
\end{equation}
where $A_{\mathrm{CW}}$ denotes the amplitude of the CW signal, $A_{\mathrm{card}}$ represents the amplitude of the modulated card’s response, $\delta(\pm f)= ({e^{j2\pi f t} + e^{-j2\pi f t}})/{2}$ representing the impulse response at the CW frequency ($f =  13.56~\mathrm{MHz}$). The term $\delta(f \pm f_s)$ denotes the card load modulation spectral components located at $f \pm f_s$, where $f_s = f/32$ or $f/28$. The button coil affects the $H(f)$, $k$, $A_{\mathrm{CW}}$ and $A_{\mathrm{card}}$ slightly since the reflected impedance in the reader coil changes its frequency response. Also, the button coil at different regions creates different magnetic coupling with the reader and card coils, which makes the PIN digit recognizable.  

\begin{table}[t]
\caption{Simulation Parameters}
\centering
\renewcommand{\arraystretch}{1.15}
\setlength{\tabcolsep}{6pt}
\begin{tabular}{c c | c c}
\hline
$\mu_0$ & $4\pi\times10^{-7}\ \mathrm{H/m}$ & 
$\omega$ & $8.52\times10^{7}\ \mathrm{rad/s}$ \\ 
\hline
$f_c$ & $13.56\ \mathrm{MHz}$ & 
$a_1$,$a_2$  &$0.0415~\mathrm{m}$, $0.0308~\mathrm{m}$\\ 
\hline
$Z_0$ & $0.1\ \Omega$ & 
$a_{p}$ & $0.0102~\mathrm{m}$ \\ 
\hline
$n_1,n_2,n_{p}$ & $3,3,2$ & 
$N$ & $9$ \\ 
\hline
$L_1$,$L_2$ & $13.8\ \mathrm{nH}$, $15.2\ \mathrm{nH}$& 
$R_{p}$ & $1.5\ \Omega$\\ 
\hline
$R_1$, $R_2$  & $1.0\ \Omega$, $0.35\ \Omega$ &
$v_1$ & $1.0~\mathrm{V}$\\
\hline
\end{tabular}
\label{tab:sim_params}
\end{table}

\begin{figure}[t]
    \centering
    \includegraphics[width=0.6\linewidth]{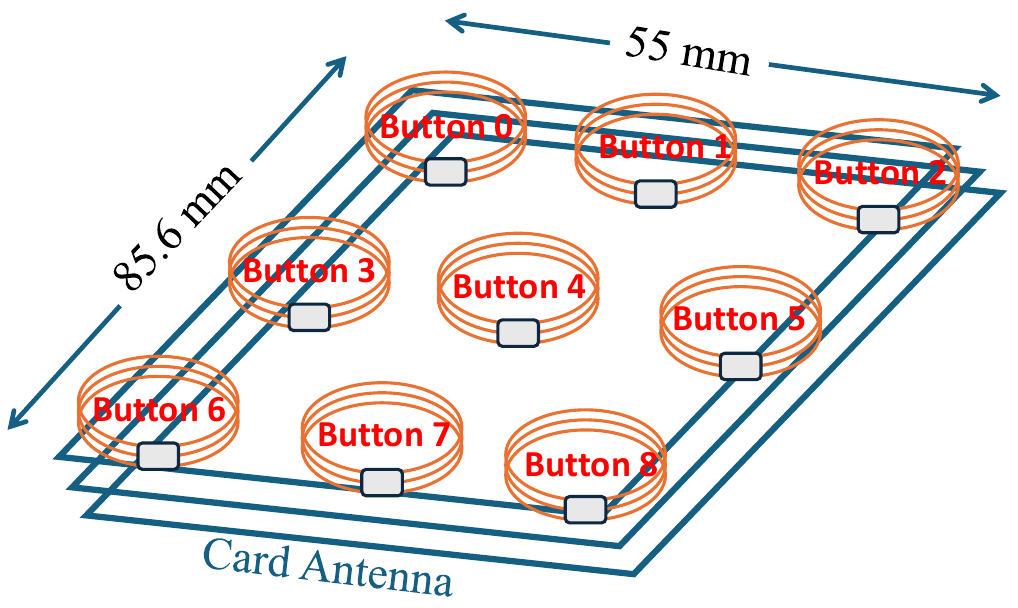}
    \caption{Buttons predefined spatial positions on an NFC card.}
    \label{fig:pad_pos_on_tag}
    \vspace{-12pt}
\end{figure}

\begin{figure}[t]
    \centering
    \includegraphics[width=0.85\linewidth]{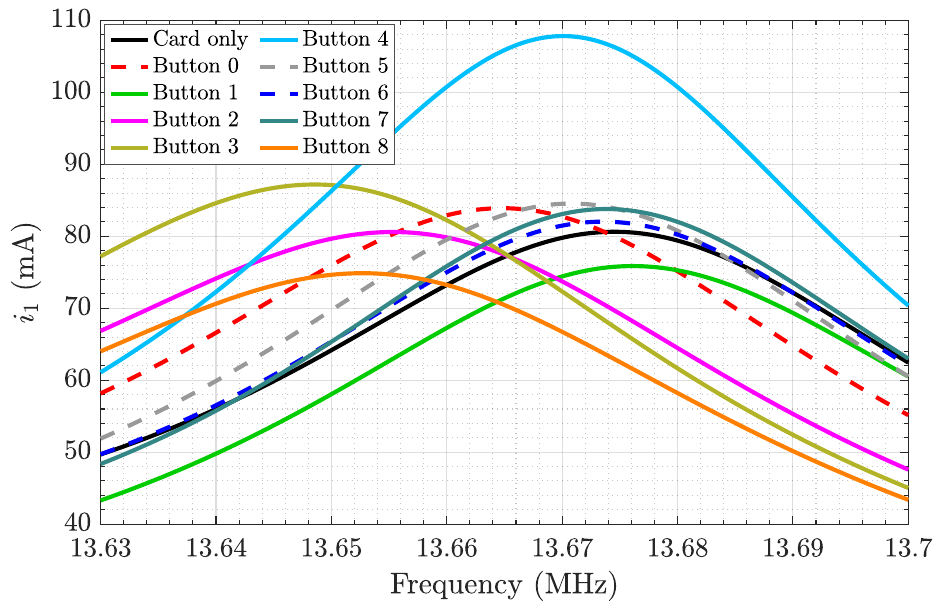}
    \caption{Effect of button coil-induced coupling on reader current variation.}
    \label{fig:ir_pad}
    \vspace{-10pt} 
\end{figure}

To evaluate the impact of the button coil, we conduct the following simulation by considering a three-coil system, which consists of the reader antenna, the card antenna, and a single button coil that is activated at a PIN digit location at a time. The NFC card surface is divided into 9 equal regions arranged in a $3\times 3$ grid that covers approximately $75\%$ of the card area, as illustrated in Fig.~\ref{fig:pad_pos_on_tag}. The centers of these regions define the locations of each virtual button, i.e., Button~0 through Button~8. The corresponding numerical simulation parameters of the system are provided in Table~\ref{tab:sim_params}. In each simulation run, a button coil is instantiated and placed at the center of a selected region, while the reader and card coils remain fixed. In the analytical model, all coils are modeled as circular loops so that their self- and mutual inductances can be efficiently computed using Bessel integral-based formulations, and the reader coil is offset toward Button~0 by 30.28\,mm and 12.47\,mm along the card width and height directions respectively, with a total in-plane offset of 32.76\,mm from card center. Fig.~\ref{fig:ir_pad} presents the variation in reader current when button coils at different regions are coupled to the reader and card antennas. As described by the reader current relation model, the total current \(i_1\) consists of the nominal (i.e., card-only) term plus a button-dependent perturbation arising from each virtual button's induced coupling reflected impedance. Each button instance modifies the effective input impedance seen by the reader, yielding measurable deviations in \(i_1\) near resonance.

\begin{figure}[t]
    \centering
    \subfigure[Simulation configuration of the predefined button locations on the card.]{
        \includegraphics[width=0.45\textwidth]{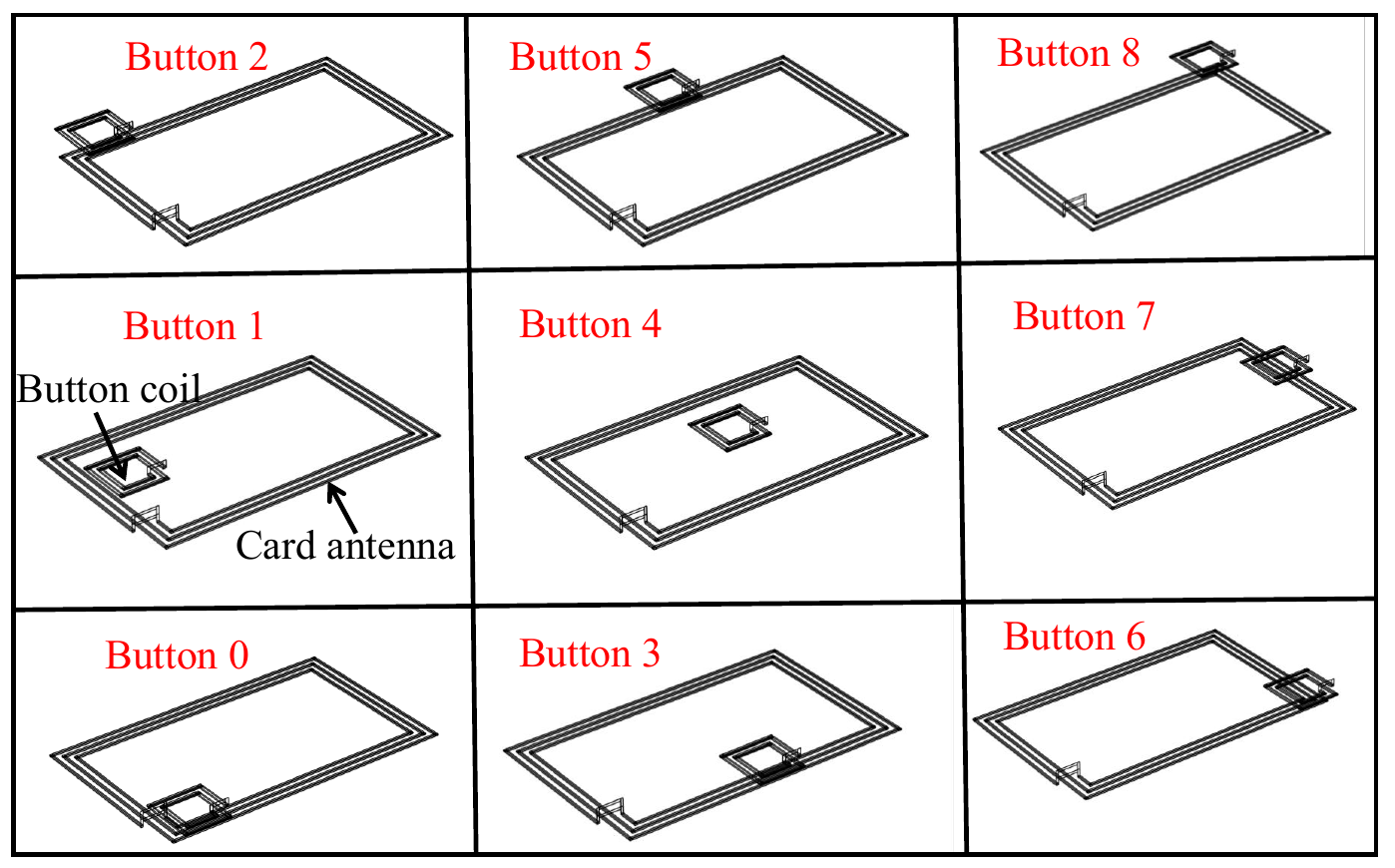}
        \label{fig:pad_array}
    }
    \hfill
    \subfigure[3D simulation model.]{
        \includegraphics[width=0.37\textwidth]{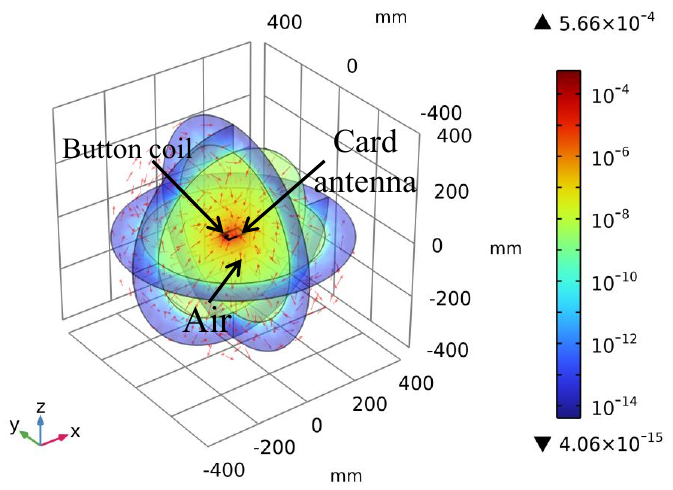}
        \label{fig:pad_sim}
    }
    \caption{COMSOL Multiphysics simulation of (a) the spatial arrangement of the virtual button on the card antenna and (b) the visualized magnetic field distribution of button–card coupling (unit: A/m).}
    \label{fig:comsol_array}
    \vspace{-10pt}
\end{figure}

\begin{figure}[t]
\hspace{-4mm}
    \centering
    \includegraphics[width=0.98\linewidth]{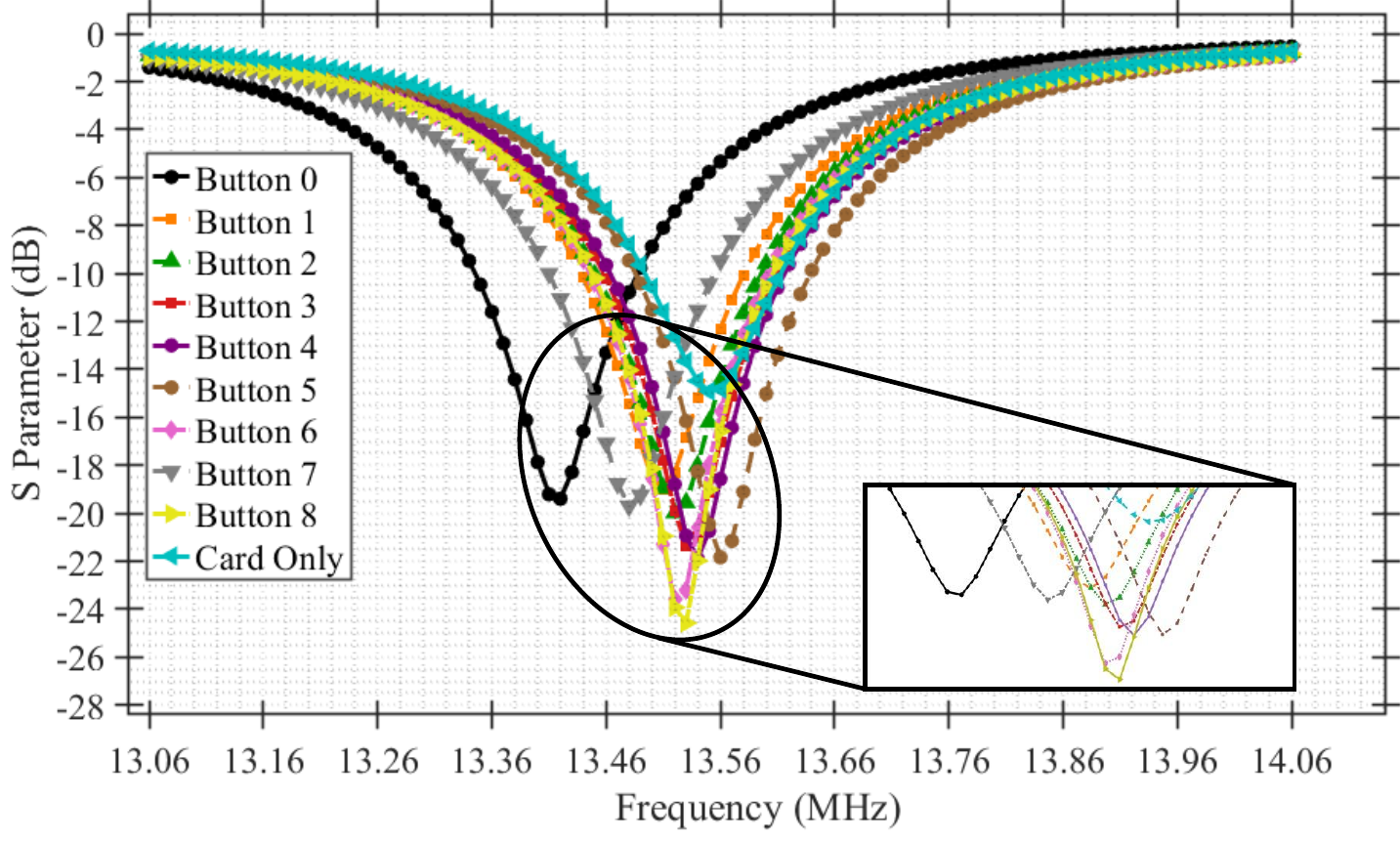}
    \caption{Impact of button coil on the reflection coefficient.}
    \label{fig:s_pam}
    \vspace{-16pt}
\end{figure}
In addition to the analytical model, we evaluate the impact of button coil in different regions close to a three-turn rectangular card antenna in COMSOL Multiphysics using the RF Module. The card antenna is actively excited, while a button coil is placed at one of the predefined virtual PIN regions and modeled using a lumped port at $13.56~\mathrm{MHz}$. Figures~\ref{fig:pad_array}–\ref{fig:pad_sim} show the virtual button positions and the resulting magnetic field distribution. The excited card antenna produces a confined near field, with each button position introducing a perturbation due to secondary coupling. As shown in Fig.~\ref{fig:s_pam}, these perturbations cause distinct variations in the reflection coefficient (S11 parameter) response without detuning the card’s primary resonance within the $13.06$–$14.06~\mathrm{MHz}$ activation band. This also validates the feasibility of recognizing the button coil location using RF signals.


\section{Prototype Design, Data Acquisition, and Virtual PIN pad Digit Recognition}
\label{Data collect}
In this section, we develop a prototype to collect data and validate the proposed CADRE for virtual PIN pad digit recognition. First, we describe the prototype system development and procedures used to collect and preprocess the NFC PIN pad dataset. Then, we introduce the domain-adversarial neural network training which is used to obtain a domain-invariant RF feature embedding. Finally, we present the PIN pad digit recognition for unseen NFC cards using Mahalanobis distance metrics. All communication follows the ISO/IEC~15693 standard \cite{iso15a}, which governs interactions between vicinity integrated circuit cards (VICCs) and vicinity coupling devices (VCDs). In this work, the VICC is referred to as the card, the VCD as the reader, and each predefined press location on the card surface as a button.

\subsection{Data Collection}

A customized ISO/IEC~15693 compliant testbed is designed and used to collect the dataset. The reader side consists of a USRP N210 connected to two FEIG ISC.ANT340/240-A loop antennas in a compact bistatic layout, with one antenna used for transmission and the other one for reception. The USRP transmits downlink commands with 100\% amplitude-shift keying (ASK) using the standard 1-out-of-4 pulse-position coding.
To ensure clarity and reproducibility, Table~\ref{tab:inventory} summarizes the exact ISO/IEC~15693 inventory request format used during data collection. 

\begin{table}[t]
\centering
\caption{ISO/IEC~15693 Inventory Request Format}
\label{tab:inventory}
\renewcommand{\arraystretch}{1.1}
\setlength{\tabcolsep}{6pt}
\begin{tabular}{l c c}
\hline
\textbf{Field} & \textbf{Bits} & \textbf{Example} \\
\hline
Start of Frame (SOF) & ---  & --- \\
Flags               & 8    & 0x02 \\
Command Code        & 8    & 0x21 \\
Optional AFI        & 0  & ---  \\
Mask Length         & 8    & 0x08 \\
CRC                 & 16   & 0xB4C3 \\
End of Frame (EOF)  & ---  & ---  \\
\hline
\end{tabular}
\vspace{-5pt}
\end{table}

A button coil is placed at different designated locations on a commercial NXP ICODE SLIX (ISO/IEC 15693 SL2S2002) 13.56~MHz card \cite{nxpicode2018}, which has the standard ID-card size of $85.6~\mathrm{mm}\times 55~\mathrm{mm}$. The nine button locations correspond to a uniformly distributed $3\times 3$ grid centered on the card antenna area, which matches the virtual button modeled in Section~\ref{system_model}. This configuration matches the commercial card geometry shown in Fig.~\ref{fig:pad_pos_on_tag}. Taking the geometric center of the card as the origin, the button coordinates are \(x \in \{-27.5,0,+27.5\}\mathrm{mm}\) and \(y \in \{+18,0, -18\}\mathrm{mm}\). This results in the button centers as follows, Button~0: \((-27.5, +18)\mathrm{mm}\), Button~1: \((0, +18)\mathrm{mm}\), Button~2: \((+27.5, +18)\mathrm{mm}\), Button~3: \((-27.5, 0)\mathrm{mm}\), Button~4: \((0, 0)\mathrm{mm}\), Button~5: \((+27.5, 0)\mathrm{mm}\), Button~6: \((-27.5, -18)\mathrm{mm}\), 
Button~7: \((0, -18)\mathrm{mm}\), and Button~8: \((+27.5, -18)\mathrm{mm}\).
 For each button position, the coil is carefully aligned using markers to ensure reproducible placement. For the uplink, the card responds in the one-subcarrier low-data-rate mode ($423.75~\mathrm{kHz}$ subcarrier, $6.62~\mathrm{kb/s}$ data rate), selected via the Flags field in the ISO/IEC~15693 inventory request in Table~\ref{tab:inventory}. This mode is chosen because its long symbols and narrowband tone yield high signal-to-noise (SNR) ratios and stable, reproducible measurements~\cite{Sarpong2025ModelAgnosticUQ}. In our configuration, the mask length is set to zero and does not include the AFI field. This yields a 48-bit inventory command. Only the Flags and CRC codes vary across commands, where Flags indicate the response mode and the CRC codes are updated accordingly. Although the dataset is collected exclusively in this mode, the acquisition system is modality-agnostic and supports all ISO/IEC 15693 response modes (including one-subcarrier low/high-data-rate and two-subcarrier low/high-data-rate)~\cite{iso15a}.

\begin{figure}[t]
  \centering
  \hspace*{3mm}
  \begin{minipage}[t]{0.85\columnwidth}
    \centering
      \includegraphics[width=\linewidth]{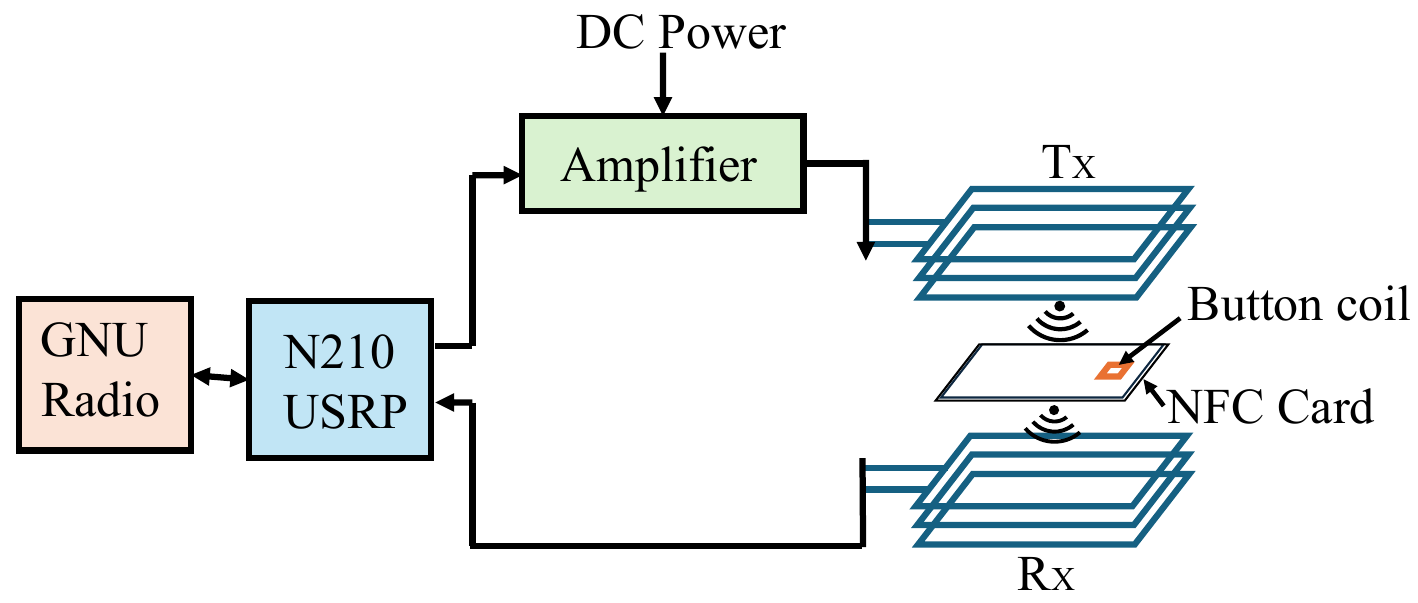}
    \captionof{figure}{Design prototype for data collection.}
    \label{fig:schemematic}
  \end{minipage}\hfill\vspace{10pt}
  \begin{minipage}[t]{0.87\columnwidth}
    \centering
    \includegraphics[width=\linewidth]{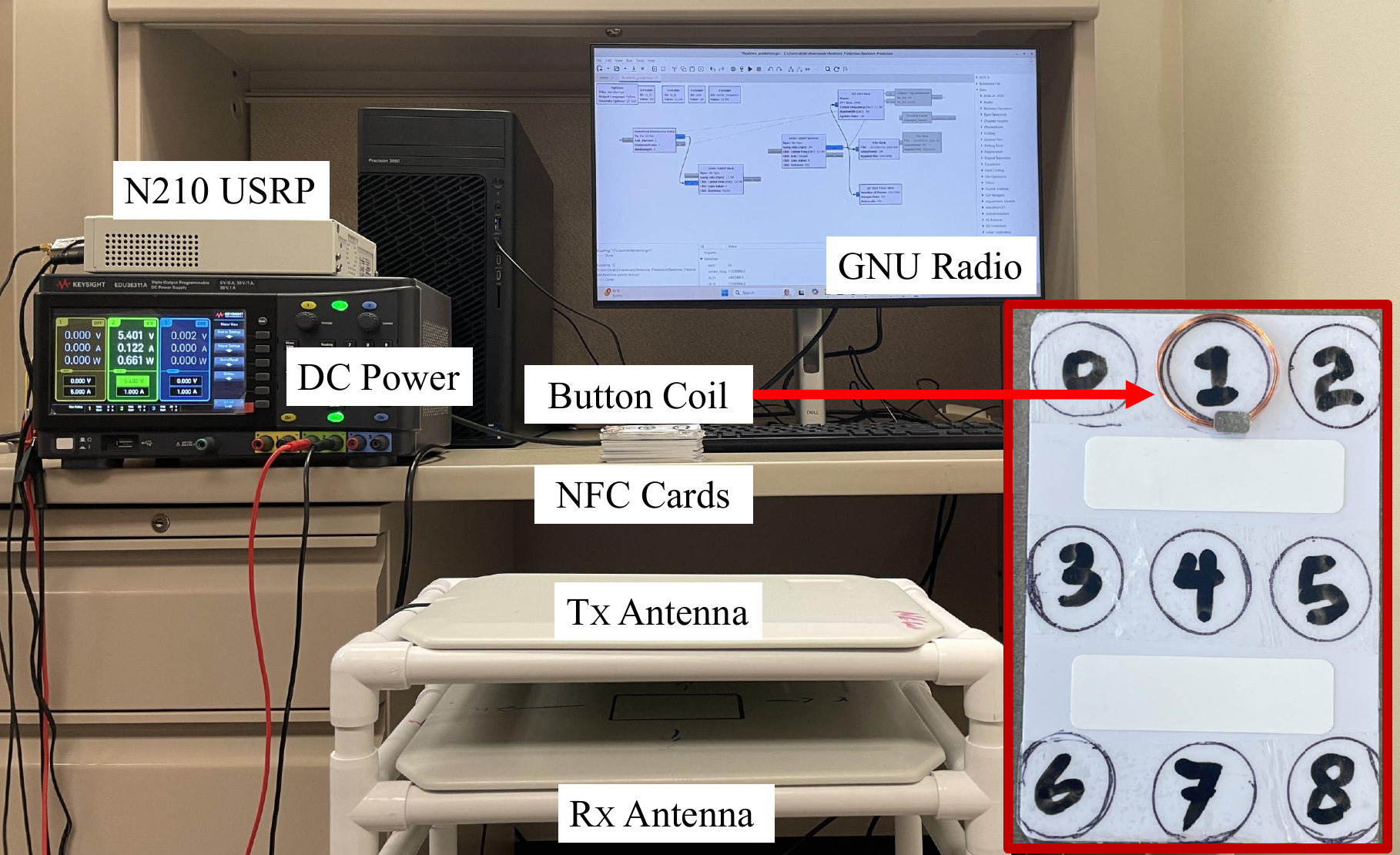}
    \captionof{figure}{Data collection experiment setup.}
    \label{fig:testbed}
  \end{minipage}
  \vspace{-15pt}
\end{figure}

The software interface is implemented in GNU Radio with two customized processing blocks for controlling the Ettus N210 USRP. We first generate ISO/IEC 15693 reader frames at a $12.5~\mathrm{M}$samples/second, including start-of-frame, command fields, and end-of-frame, followed by an idle slot and reset interval to emulate the pause before the card reply. Then, we receive complex baseband samples at $2~\mathrm{M}$samples/second using the known command duration to isolate the expected reply window. The start of the response is detected using a correlation-based trigger by matching the received card response signal with a reference signal. After detecting the start, a fixed segment of the response is extracted for recognition. In particular, the first 8 response bits are used, as they are constant across all cards and independent of payload data. Also, we remove slow current drifts and normalize the received signal to ensure consistency across sessions. To compensate for the limited transmission power of the USRP and guarantee reliable activation of ISO/IEC 15693 cards, a Mini-Circuits ZX60-100VHX+ RF amplifier is added to the transmit chain.

\begin{figure}[t]
    \centering
    \includegraphics[width=0.9\linewidth]{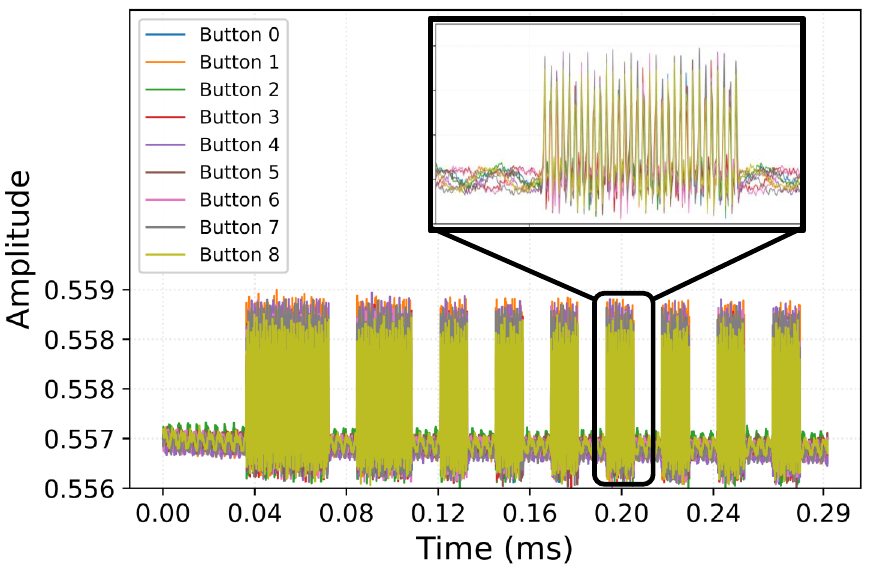}
    \caption{Response signals across 9 button positions on a card.}
    \label{fig:pad_press_response}
    \vspace{-15pt}
\end{figure}

For each button coordinate, we place the button coil at the specified location, while the reader sends repeated inventory commands that specify one-subcarrier low-data-rate responses. To capture variability, recordings are made in five different orientations of the card relative to the reader antennas. In each orientation, exactly $100$ valid responses are accepted, with any unsynchronized traces replaced, resulting in $500$ responses per button. With $9$ buttons per card, this yields $4{,}500$ responses per card. In total, button-press responses were collected from $16$ NFC cards, producing an overall dataset of $72{,}000$ response samples. An illustration of the experimental prototype for the data acquisition system is shown in Fig.~\ref{fig:schemematic}, and the associated implementation is given in Fig.~\ref{fig:testbed}. Figure~\ref{fig:pad_press_response} visualizes the per-button responses, which exhibit subtle yet consistent distinctions across button locations.

\subsection{Domain-Adversarial Training}
We use a Domain-Adversarial Neural Network (DANN) to handle cross-card PIN pad digit recognition. In practice, we can only obtain limited data from cards we have, and the DANN is used to extend the solution to unseen cards (the same type cards). In this work, we refer to a source domain as labeled button press signals collected from known NFC cards, while the target domain refers to unlabeled button press signals collected from new cards. The model learns features that separate button press signal classes, while ignoring card-specific RF fingerprints. This is achieved by combining a supervised cross-entropy loss on labeled source data with an adversarial domain loss using a Gradient Reversal Layer (GRL). The encoder produces card-invariant yet class-discriminative representations \cite{Zhang2023UnsupervisedDA,9730081}.

\begin{figure}[t]
    \centering
    \includegraphics[width=\linewidth]{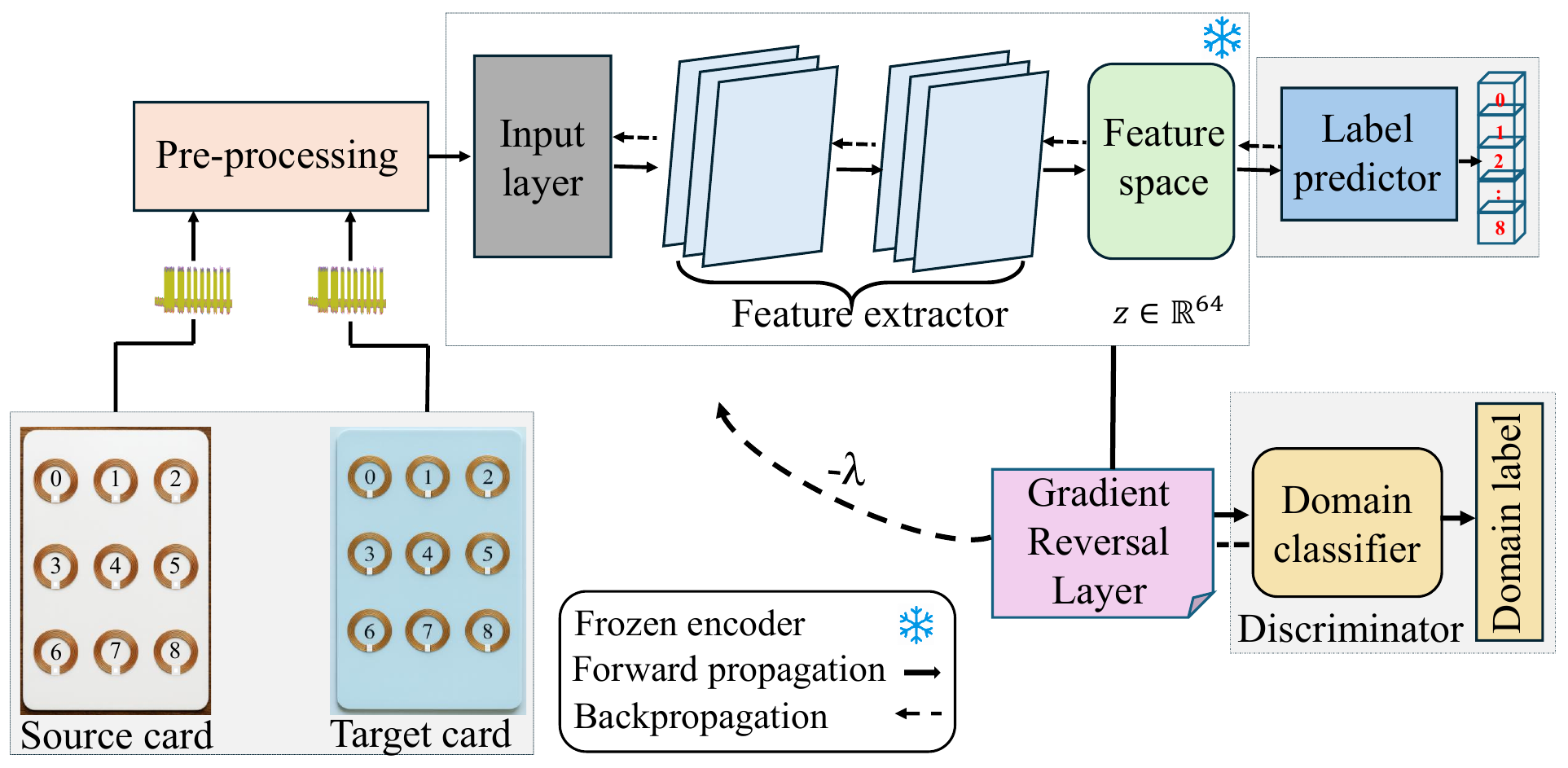}
    \caption{Adversarial domain adaptation architecture with shared encoder and domain discriminator.}
    \label{fig:Workflow}
    \vspace{-15pt}
\end{figure}

We use three loss items jointly: a source classification loss that makes button classes separable, an adversarial domain loss that removes domain-specific cues, and a target entropy term that sharpens unlabeled signal samples. Overall, the classification loss encourages the encoder to keep button embeddings distinct, whereas the adversarial term encourages the encoder to hide any card-specific information. The training architecture in Fig.~\ref{fig:Workflow} combines a temporal encoder, a PIN digit class head, and a domain (card) head linked by a GRL.

The encoder maps each NFC button signal sample to a 64-dimensional embedding. It first applies two one-dimensional (1D) convolutional blocks with 7-tap kernels. The first block uses 32 filters, and the second uses 64 filters. Each convolutional block is followed by batch normalization and max pooling. These layers capture local signal patterns and reduce the temporal resolution. The convolutional layers learn short-term amplitude and slope variations introduced by button presses. Next, a 64-unit Long Short-Term Memory (LSTM) layer is applied. This layer models longer-term temporal structures in the signal. The LSTM captures how the signal evolves over time. This temporal evolution provides unique information for each button. After the recurrent stage, global average pooling reduces the temporal dimension. A dense projection then produces a 64-dimensional feature vector.

The classifier is a multilayer perceptron (MLP) that first applies a fully connected layer with 64 units and ReLU activation, followed by batch normalization, and then a fully connected layer with 9 outputs and a softmax activation. It is trained with cross-entropy loss on the labeled source data to produce class posteriors for the 9 NFC button press signals. The domain classifier (discriminator) is another MLP that operates on the 64-dimensional encoder embedding. It takes the 64-dimensional feature vector as input and passes it through a fully connected layer with 128 units and ReLU activation. A dropout layer with rate 0.5 is then applied for regularization. Next, a fully connected layer with 64 units and ReLU activation further processes the features. The final layer is a single-unit output with a sigmoid activation that predicts the probability that an embedding originates from the source domain rather than the target domain. A GRL is placed between the encoder and the discriminator. It forwards the features unchanged but multiplies the backpropagated gradient by a weight factor $-\lambda$, where $\lambda$ is a positive weight. As a result, the discriminator is trained to distinguish source from target features, while the encoder is encouraged to produce domain-invariant representations that make this discrimination difficult.

We represent the labeled source mini-batch as $\mathcal{B}_s = \{(x_r^s, y_r^s)\}_{r=1}^{m_s},~y_r^s \in \{0, \dots, C-1\}$,
and the unlabeled target mini-batch as $\mathcal{B}_t = \{x_w^t\}_{w=1}^{m_t}$, where $m_s$ and $m_t$ denote the number of source and target samples in a mini-batch, respectively. The dataset includes many mini-batches.
These two types of mini-batches serve complementary roles in training: $\mathcal{B}_s$ defines the supervised class, while $\mathcal{B}_t$ provides the signal required to enforce domain invariance across different NFC cards.

The encoder $f_{\theta} : \mathcal{X} \rightarrow \mathbb{R}^d$ maps each input signal to a $d$-dimensional embedding, with $d = 64$ in all experiments.
The classifier $g_{\phi} : \mathbb{R}^d \rightarrow \mathbb{R}^C$ maps embeddings to digit button-class logits, while the discriminator $h_{\psi} : \mathbb{R}^d \rightarrow \mathbb{R}$ predicts the domain class.
The class posterior is given by $\hat{p}(y | x) = \mathrm{softmax}\!\big(g_{\phi}(f_{\theta}(x))\big),$
and the domain posterior is $\hat{p}(b = 1| z) = \sigma\!\big(h_{\psi}(z)\big)$,
where $\sigma(\cdot)$ denotes the logistic sigmoid.
The class posterior measures the likelihood that a signal corresponds to a specific button class, while the domain posterior reflects the confidence that an embedding originates from the target card domain.

The supervised classification loss over a labeled source mini-batch is defined as
\begin{equation}
\mathcal{L}_{\mathrm{cls}} =
\frac{1}{m_s} \sum_{r=1}^{m_s}
\left[-\log \hat{p}_{y_r^s}\!\big(y = y_r^s \mid x_r^s\big)\right].
\end{equation}
This loss enforces correct button classification and promotes separation between button-specific embeddings in the latent space. Moreover, the adversarial domain loss computed over a mixed source-target mini-batch is given by
\begin{align}
\mathcal{L}_{\mathrm{dom}}
&= \frac{1}{m_s} \sum_{r=1}^{m_s}
\left[-\log\!\big(1 - \hat{p}(b = 1 \mid f_{\theta}(x_r^s))\big)\right]
\notag\\
&\quad + \frac{1}{m_t} \sum_{w=1}^{m_t}
\left[-\log \hat{p}(b = 1 \mid f_{\theta}(x_w^t))\right].
\end{align}
This loss trains the discriminator to distinguish source and target embeddings, while the GRL ensures that the encoder receives a sign-reversed gradient. Consequently, the encoder is driven to suppress card-specific variations while preserving class-discriminative structure.

To further encourage confident predictions on unlabeled target data, an entropy regularization term is introduced:
\begin{equation}
\mathcal{L}_{\mathrm{ent}} =
\frac{1}{m_t} \sum_{w=1}^{m_t}
\left[-\sum_{c=0}^{C-1}
\hat{p}_c\!\big(y = c \mid x_w^t\big)
\log \hat{p}_c\!\big(y = c \mid x_w^t\big)\right].
\end{equation}
This term penalizes high-entropy predictions, thereby implicitly pulling target embeddings toward the nearest class clusters defined by the labeled source data.

The encoder-classifier parameters $(\theta, \phi)$ and the discriminator parameters $\psi$ are optimized according to
\begin{equation}
\min_{\theta, \phi}\; \mathcal{L}_{\mathrm{cls}} + \gamma_t \mathcal{L}_{\mathrm{ent}} - \lambda \mathcal{L}_{\mathrm{dom}},
\qquad
\min_{\psi}\; \mathcal{L}_{\mathrm{dom}} .
\label{optimization}
\end{equation}
The parameter $\gamma_t$ controls the contribution of the entropy regularization term, while $\lambda$ scales the strength of the adversarial signal and is gradually increased during early training epochs. All loss terms are evaluated within each mini-batch, and training is carried out using mini-batch stochastic gradient descent. For each joint mini-batch $(\mathcal{B}_s, \mathcal{B}_t)$, the classification, domain, and entropy losses are computed. The discriminator parameters $\psi$ are updated to minimize the domain loss $\mathcal{L}_{\mathrm{dom}}$, which improves separation between source and target embeddings. Also, the encoder-classifier parameters $(\theta,\phi)$ are updated according to Eq.~\eqref{optimization}. The GRL inverts the gradient of $\mathcal{L}_{\mathrm{dom}}$ before it reaches the encoder, such that the encoder is optimized to reduce classification and entropy losses while simultaneously learning representations that are insensitive to domain differences. As a result, the discriminator and encoder are updated with complementary objectives within the same training loop. Thus, the domain invariance and class discrimination are learned jointly rather than sequentially. Balanced mini-batches with $m_s \approx m_t$ are used throughout training.

At inference time, we have a target button signal input \(x^{t}\), and we do not know which button or card it is associated with. The recognition uses only the encoder \(f_{\theta}\) and the classifier \(g_{\phi}\):
\begin{equation}
\hat y=\arg\max_{c\in\{0,\dots,C-1\}}
\hat p_c\!\big(y{=}c\mid x^t\big).
\end{equation}
Here, \(x^t\) is a target-domain input at test time, \(\hat y\) is the predicted button class, and \(\hat p_c(y{=}c| x^t)\) is generated by the deep learning model's output $\mathrm{softmax}(g_{\phi}(f_{\theta}(x^t)))$. To ensure the effectiveness of evaluation, target labels are strictly held out and not used for adaptation, hyperparameter selection, early stopping, or calibration. They are reserved exclusively for evaluation after adversarial domain adaptation.

\begin{figure}[t]
    \centering
    \subfigure[without domain adaptation]{
        \includegraphics[width=0.225\textwidth]{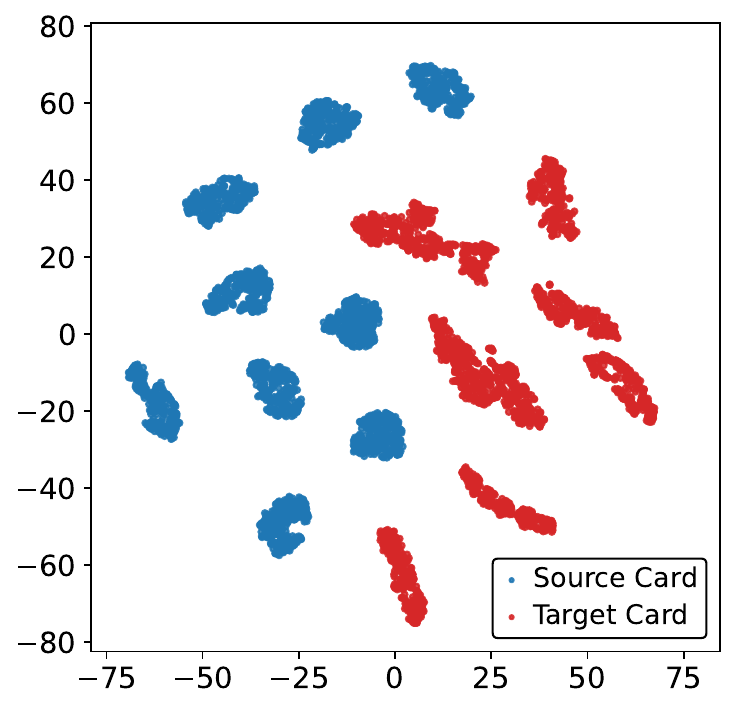}
        \label{fig:tsne-no-da}
    }
    \subfigure[with domain adaptation]{
        \includegraphics[width=0.225\textwidth]{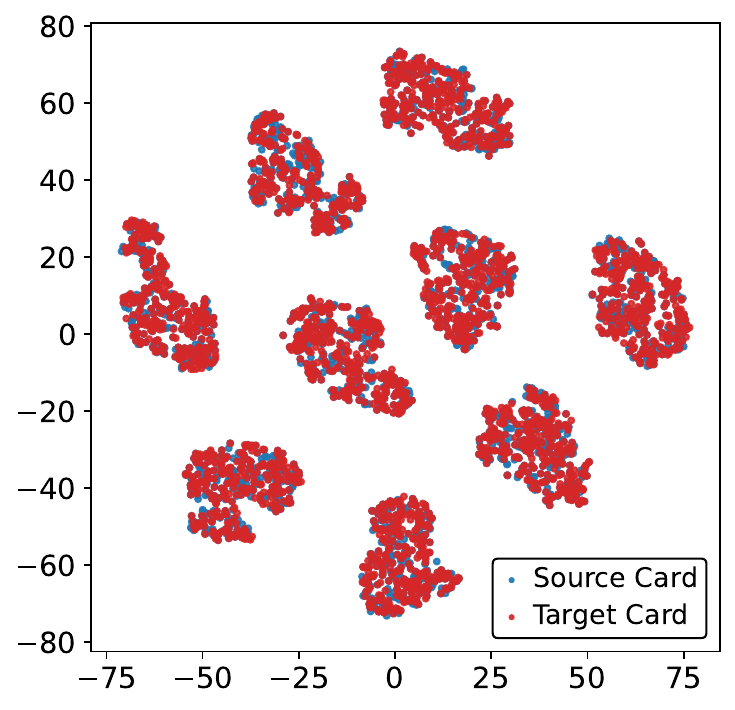}
        \label{fig:tsne-da}
    }
    \caption{t-SNE visualization of encoder embeddings with and without domain adaptation for source and target NFC card button classes.}
    \label{fig:tsne-panels}
    \vspace{-10pt}
\end{figure}

\begin{figure}[t]
    \centering
    \subfigure[Source.]{
        \includegraphics[width=0.226\textwidth]{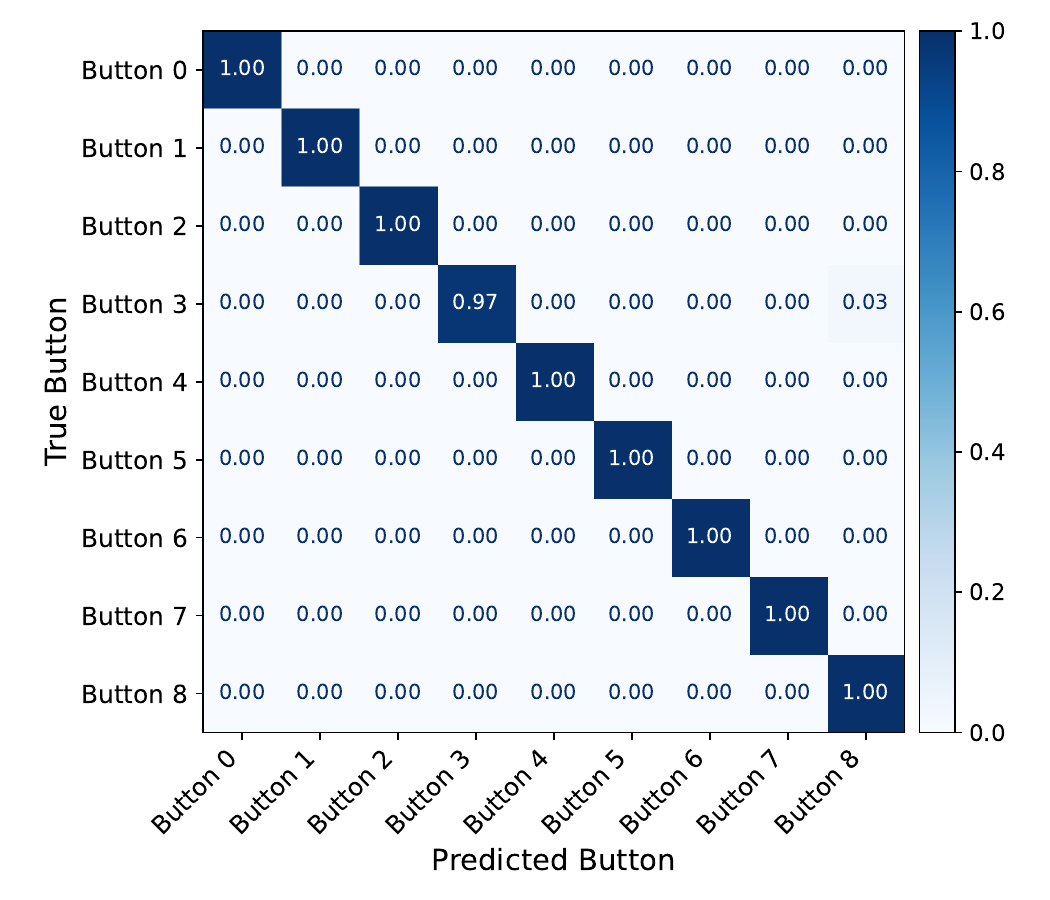}
        \label{fig:cm_source}
    }
    \subfigure[Target.]{
        \includegraphics[width=0.226\textwidth]{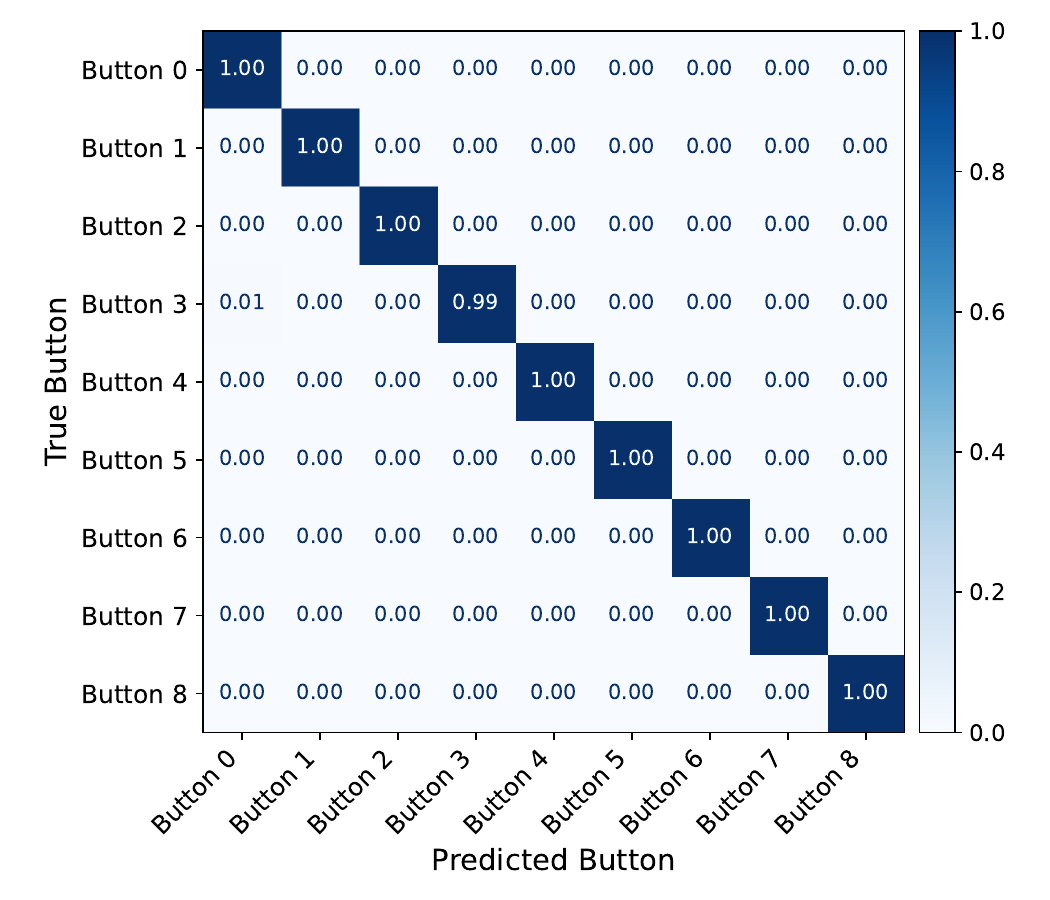}
        \label{fig:cm_target}
    }
    \caption{Confusion matrices for button class prediction using the adapted encoder.}
    \label{fig:cm_source_target}
    \vspace{-15pt}
\end{figure}

To illustrate the effect of the adaptation objective, Fig.~\ref{fig:tsne-panels} shows a t-SNE visualization. First, without domain adaptation, source and target features are clustered predominantly by domain. Then, after adaptation, the embeddings align within shared class clusters, which indicates that the encoder successfully removes card-dependent distortions while preserving button-specific structure. To better understand how reliably the adapted encoder--classifier model generalizes, we evaluate it using two separate source--target pairings, where each pairing uses signals from 2 NFC cards for training, resulting in a total of 4 training cards across both evaluations, and the results are shown in Fig.~\ref{fig:cm_source_target}. Across both evaluations, the model achieves average source test accuracy of $99.56\% \pm 0.69\%$ and average target test accuracy of $99.72\% \pm 0.19\%$. After domain adaptation, the encoder parameters are fixed. The frozen encoder is saved and subsequently used as a feature extractor for the PIN pad input recognition.

Based on the trained model, we develop a scheme that determines whether a button-press signal truly belongs to a valid button class or corresponds to an unknown or invalid press. Although the encoder--classifier pipeline can always output a button label for an unseen input, it may still assign high confidence to predictions when the input signal lies outside the distribution of legitimate button presses. Such out-of-distribution button-press signals can therefore be confidently misclassified as valid button classes, posing risks to reliable system operation. In practice, this situation arises when users press outside the predefined button regions, generating signals that do not correspond to any valid button. Under these conditions, reliance on classifier confidence alone can lead to erroneous recognition \cite{10.5555/3618408.3618513}. 

To address this challenge, we introduce a recognition mechanism that explicitly evaluates whether an unseen button-press signal is statistically consistent with any learned button class. If the signal cannot be reliably recognized, the system requests the user to re-enter the input. This button-press signal recognition is grounded in a distance-based statistical model using the encoder’s latent features. Rather than just relying on classifier confidence, this scheme evaluates how geometrically and statistically consistent an unseen button-press signal's embedding is compared with the learned distributions of legitimate button press signal feature. At the core of this framework is the Mahalanobis distance, a covariance-aware metric that measures how far a sample lies from a reference cluster in a multivariate space. Unlike the Euclidean distance, which assumes each feature dimension is independent and equally scaled, the Mahalanobis metric incorporates correlations and unequal variances among latent dimensions. Geometrically, it defines an adaptive decision boundary that conforms to the true spread of the data rather than forming a fixed circular boundary.
\subsection{PIN Digit Button-Press Recognition}
\begin{figure}
    \centering
    \includegraphics[width=0.50\linewidth]{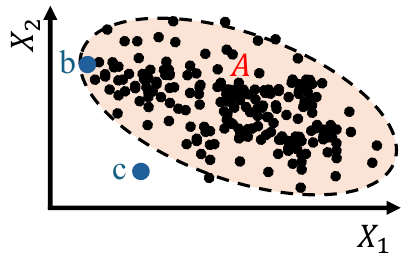}
    \caption{Illustration of Mahalanobis distance concept.}
    \label{fig:MD_llustration}
    \vspace{-16pt}
\end{figure}

In Fig.~\ref{fig:MD_llustration}, we illustrate the geometric interpretation of the Mahalanobis distance in a 2-dimensional feature space $(X_1, X_2)$. The black cluster $A$ represents latent feature samples of a legitimate button press class, which is bounded by a Mahalanobis-distance contour derived from a predefined risk level ($\alpha$). The dashed ellipse defines this acceptance boundary based on the class covariance. Samples that fall inside the ellipse have a Mahalanobis distance below the decision threshold and are accepted as legitimate. Point~$b$, although shifted away from the cluster center, remains inside the bounded region and is therefore accepted. Point~$c$ lies outside the boundary which exceeds the distance threshold, which is rejected as non-legitimate.

For each button-press signal sample, the captured response is transformed by the encoder into a latent vector \( f_{\theta}(x) \). Each button class forms a distribution with a mean and a covariance that reflect the typical spread of its latent vectors. To characterize these distributions, we compute class-wise mean vectors and a covariance matrix using calibration data. For button class \(c\), let \(D_{c}^{\text{cal}}\) contain \(N_c\) calibration samples. Its mean latent vector is computed as
\begin{equation}
    u_c = \frac{1}{N_c} \sum_{x \in D_c^{\text{cal}}} f_{\theta}(x),
\end{equation}
which represents the centroid of class \(c\) in the latent space.

Let \( z = f_{\theta}(x) = [z_1, \ldots, z_d]^\top \) represent the \(d\)-dimensional latent embedding and $(\cdot)^\top$ denote the transpose operation. The covariance matrix \(S\) across all classes is defined as \cite{Rafiq2024EfficientIO}
\begin{equation}
S =
\begin{bmatrix}
\mathrm{Var}(z_1) & \mathrm{Cov}(z_1,z_2) & \cdots & \mathrm{Cov}(z_1,z_d)\\
\mathrm{Cov}(z_2,z_1) & \mathrm{Var}(z_2) & \cdots & \mathrm{Cov}(z_2,z_d)\\
\vdots & \vdots & \ddots & \vdots\\
\mathrm{Cov}(z_d,z_1) & \mathrm{Cov}(z_d,z_2) & \cdots & \mathrm{Var}(z_d)
\end{bmatrix},
\label{eq:cov_matrix}
\end{equation}
where $\mathrm{Var}(z_m)$ denotes the variance of the $m$th latent dimension,
$\mathrm{Cov}(z_m,z_n)= \frac{1}{N_t} \sum_{c \in \mathcal{C}} \sum_{x \in D_c^{\text{cal}}}
\big(z_m - u_{c,m}\big)\big(z_n - u_{c,n}\big)$, $m,n \in \{1,\ldots,d\}$ index the latent feature dimensions, $u_{c,m}$ and $u_{c,n}$ denote the $m$th and $n$th components of the class
mean vector $u_c$, respectively, and  \(N_t = \sum_{c\in\mathcal{C}} N_c\). When \(S = I_d\), the latent dimensions are decorrelated with unit variance, and the Mahalanobis distance reduces to the Euclidean distance. Thus, \(S\) captures both the scale and the correlation structure of the latent features under PIN digit variability.

The squared Mahalanobis distance between a test button-press signal's embedding \(f_\theta(x_i)\) and class \(c\) is \cite{Lu2023DiversifyAG}
\begin{equation}
    d_i^{2}(x_i)
    = \big(f_{\theta}(x_i) - u_c\big)^{\top}
      S^{-1}
      \big(f_{\theta}(x_i) - u_c\big).
\end{equation}
The inverse covariance matrix \(S^{-1}\) rescales the space so that deviations in low variance directions contribute more strongly and those in high variance directions contribute less. A small distance indicates that the sample is statistically compatible with class \(c\), while a large distance suggests mismatch or spoofing.

To avoid explicitly computing \(S^{-1}\), which can be numerically unstable, we adopt a Cholesky factorization \(S = L L^\top\), where \(L\) is lower-triangular. This yields the equivalent and stable form \cite{10.1007/s00180-022-01277-6}:
\begin{equation}
    d_i(x_i) = \big\| L^{-1} \big(f_{\theta}(x_i) - u_c\big) \big\|_2^2.
    \label{cholesky}
\end{equation}
This interpretation shows that the Mahalanobis distance is the squared Euclidean norm of a whitened vector, where the whitening transformation \(L^{-1}\) removes correlations and normalizes variances according to the covariance structure.

After obtaining the Mahalanobis distances from each test embedding to all button class centroids, the system must determine whether the nearest class is statistically valid. To support this decision, we characterize the distribution of distances produced by non-matching button classes. For each button class $c$, we construct an empirical impostor distance set, i.e., distances between the centroid of class \(c\) and calibration samples from all other classes:
\begin{equation}
D_c^{\text{imp}}
=
\left\{
\left\|L^{-1}\big(f_{\theta}(x)-u_c\big)\right\|_2^2 :
x \in D_q^{\text{cal}},\ q \neq c
\right\},
\label{imposters_cal}
\end{equation}
where \(D_q^{\text{cal}}\) denotes the calibration datasets of all button classes other than the reference class \(c\). The class $c_i$ is obtained by selecting the class whose centroid minimizes the Mahalanobis distance to the test signal $x_i$, given by $c_i = \arg\min_{c\in\mathcal{C}} d_i(x_i).$ Specifically, after identifying the class $c_i \in \mathcal{C}$ whose centroid yields the minimum Mahalanobis distance to a test sample, the corresponding class-specific threshold is obtained as
\begin{equation}
t_{c_i}(\alpha)
=
Q_{\alpha}\!\left(D_c^{\mathrm{imp}}\right), 
\end{equation}
where ${c=c_i}$ and $Q_{\alpha}(\cdot)$ denotes the empirical $\alpha$-quantile operator. The risk level $\alpha$ specifies the tolerated fraction of impostor distances that fall below the threshold. 

Given the selected button class, the decision rule for PIN digit class acceptance is then defined as
\begin{equation}
\hat{a}(x_i,c_i;\alpha) =
\begin{cases}
1, & \text{if } d_i(x_i) \leq t_{c_i}(\alpha),\\[3pt]
0, & \text{otherwise},
\end{cases}
\end{equation}
where \(\hat{a}(x_i,c_i;\alpha)\in\{0,1\}\) denotes the acceptance decision.

The time complexity of the proposed button-press recognition pipeline is determined by the covariance modeling and distance evaluation stages. Constructing the  covariance matrix from $N_{\text{c}}$ latent vectors requires $O(N_{\text{c}} d^{2})$, whereas obtaining the whitening transform via the Cholesky factorization of the $d \times d$ covariance matrix incurs $O(d^{3})$. During inference, evaluating the Mahalanobis distance for each class involves solving a triangular system followed by a squared norm computation, giving a per button class cost of $O(d^{2})$ \cite{Shrivastava2023AnomalyDU}. The overall time complexity for recognizing a single test sample is $O(C d^{2})$, where $C$ is the number of button classes. Hence, the most time-consuming operation in our method is the Cholesky factorization with complexity $O(d^{3})$; however, this step is incurred only once during calibration, and the per sample recognition cost remains $O(C d^{2})$, making the procedure efficient for real-time NFC virtual button-press recognition tasks.


\section{Experiments and Results}
\label{Results}
In this section, we evaluate the proposed CADRE framework on unseen cards and analyze its performance using standard recognition metrics: False Acceptance Rate (FAR), False Rejection Rate (FRR), and Acceptance Rate (AR). AR is the fraction of test samples that are accepted and correctly assigned to their true button classes. Let the test set be \(X_{\text{test}} = \{x_1,\ldots,x_{N_{\text{test}}}\}\) with the corresponding true labels \(\{y_1,\ldots,y_{N_{\text{test}}}\}\). AR is given as
\begin{equation}
\mathrm{AR}
=\frac{1}{N_{\text{test}}}\sum_{i=1}^{N_{\text{test}}}
\mathbb{I}\!\left[c_i = y_i \;\text{and}\; \hat{a}(x_i,c_i;\alpha)=1\right],
\label{AR}
\end{equation}
where \( \mathbb{I}(\cdot) \) is the indicator function that returns 1 if, for each test sample \(x_i\), the scheme outputs a correct class prediction \(c_i = y_i\) and an acceptance decision \(\hat{a}(x_i,c_i;\alpha)=1\). For the subsequent recognition stage, we withheld 10\% of the source card data for calibration. This split was not used for training, adaptation, or evaluation. The target card dataset was used entirely for evaluation. To evaluate generalization, we use a dataset from 12 unseen NXP ICODE SLIX (ISO/IEC~15693 SL2S2002) cards, with each card providing 500 responses per button across all $9$ virtual buttons, resulting in 54{,}000 test samples.

\begin{figure}[t]
    \centering
    \includegraphics[width=0.75\linewidth]{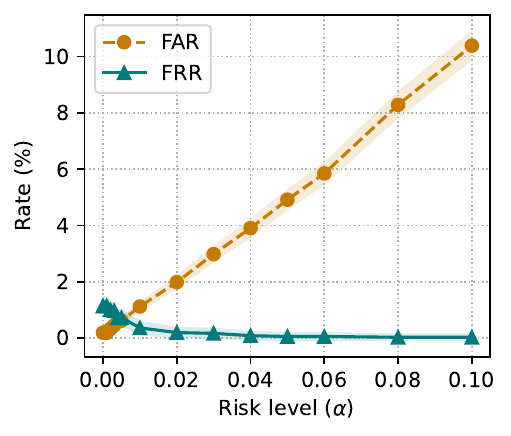}
    \caption{Impact of risk levels ($\alpha$) on FAR and FRR.}
    \label{fig:pad_far_frr_alpha}
    \vspace{-15pt}
\end{figure}

\begin{figure}[t]
    \centering
    \includegraphics[width=\linewidth]{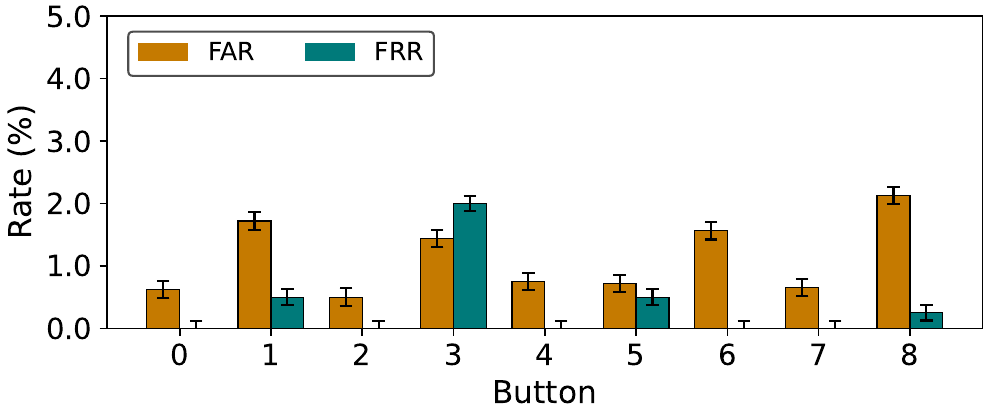}
    \caption{Average per-button FAR and FRR for unseen cards.}
    \label{fig:pad_far_frr}
    \vspace{-10pt}
\end{figure}
In Fig.\ref{fig:pad_far_frr}, the average per-button FAR and FRR are shown. FAR measures the percentage of incorrect responses accepted as genuine, whereas FRR measures the percentage of genuine responses incorrectly rejected. Error rates remain below 2.00\% for most buttons, with slight increases on Button~3 and Button~8. Overall, the low and consistent error rates indicate that the learned thresholds transfer effectively to new card hardware and maintain balanced sensitivity and specificity across all the buttons.
\begin{figure}[t]
    \centering
    \includegraphics[width=0.76\linewidth]{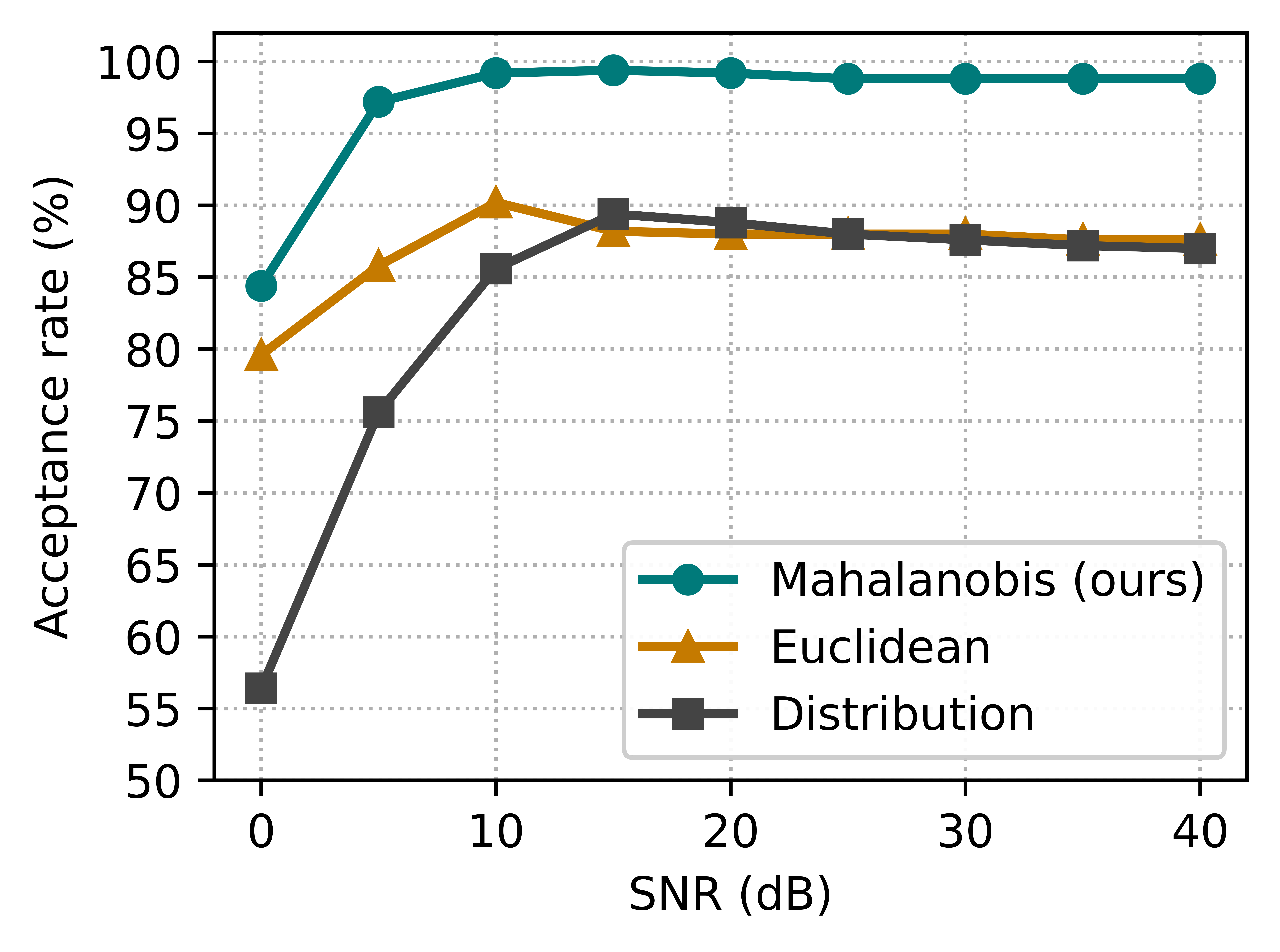}
    \caption{Recognition performance using different methods.}
    \label{fig:compare_to_baselines}
    \vspace{-14pt}
\end{figure}

The impact of the acceptance risk on the tradeoff between FAR and FRR is evaluated by varying $\alpha$ over $\alpha \in \{0.01,0.02,0.04,0.06,0.08,0.10\}$. As shown in Fig.~\ref{fig:pad_far_frr_alpha}, FAR increases gradually with $\alpha$, while FRR remains consistently low. Based on this empirical analysis, we select ${\alpha =0.025}$, which provides an optimum operating point with low error rates and stable generalization across all our experiments.

\begin{table}[t]
\centering
\caption{Recognition Performance Comparison}
\begin{tabular}{lccc}
\hline
\textbf{Method} & \textbf{FAR (\%)} & \textbf{FRR (\%)} & \textbf{AR (\%)} \\
\hline
Euclidean              & 13.25 & 12.60 & 87.17 \\
Distribution           & 17.18 & 16.33 & 82.84 \\
Mahalanobis (ours)     & \textbf{1.84} & \textbf{1.63} & \textbf{98.20} \\
\hline
\end{tabular}
\label{tab:recognition_methods}
\vspace{-15pt}
\end{table}

Next, we evaluate the system under noise by injecting AWGN with controlled SNR levels directly into the raw button-press signals before normalization and encoder processing. Three button-press recognition strategies are compared: Euclidean distance to the class centroid, a Distribution-based likelihood method that models the latent features using class-conditional probability distributions that share the same feature spread and correlation across all classes \cite{Venkataramanan2023GaussianLR}, and the proposed Mahalanobis distance-based approach. The experiments shown in Fig.~\ref{fig:compare_to_baselines} compare the AR trends of all methods under increasing noise severity. The Euclidean and Distribution-based approaches show significant performance drops as SNR decreases, while the Mahalanobis approach maintains consistently higher AR across all noise levels. The results reflect the behavior of each method when operating on degraded embeddings produced under identical noise conditions.

As summarized in Table~\ref{tab:recognition_methods}, the Mahalanobis approach significantly outperforms the Euclidean and Distribution  baselines, reducing both FAR and FRR and achieving an AR of 98.20\%. This improvement is consistent with the trends observed in Fig.~\ref{fig:compare_to_baselines}, particularly under low-SNR conditions, where the covariance aware geometric modeling of the Mahalanobis scheme provides superior robustness.

\section{Conclusion}
\label{Conclude}
This paper introduces a card-invariant virtual PIN pad system on passive NFC cards that combines domain adaptation with statistical distance-based solution to achieve consistent recognition of button responses across different ISO/IEC 15693 cards. By placing passive resonant coils at predefined positions on the card, the reader-card-coil interaction produces distinct analog response patterns for each virtual button. A deep learning model is used to learn representations that capture button-specific characteristics while suppressing variations caused by card hardware. The proposed CADRE (Card-Agnostic Domain-Aligned RF Embedding) framework enables accurate PIN digit button-press recognition even on previously unseen cards. A Mahalanobis-based decision mechanism keeps FAR and FRR low. CADRE delivers a fast, passive, and reliable button-press recognition approach for passive NFC PIN pad interactions.

\section*{Acknowledgment}
This work was supported in part by the National Science Foundation under grant no. CNS-2341846 and CNS-2310856.

\bibliographystyle{IEEEtran}
\fontsize{8.5pt}{8.5pt}\selectfont
\bibliography{ref}

\end{document}